\documentclass[referee]{aa} 
\usepackage{graphicx}
\usepackage{natbib}
\usepackage{amssymb}
\usepackage{relsize}
\usepackage{txfonts}
\usepackage{rotate}
\usepackage{epsf}
\newcommand{\Alfven}{$\rm Alfv\acute{e}n$}

\newcommand{\D}{\partial}
\newcommand{\DD}{\frac}

\newcommand{\beq}{\begin{equation}}
\newcommand{\eeq}{\end{equation}}
\newcommand{\ben}{\begin{enumerate}}
\newcommand{\een}{\end{enumerate}}
\newcommand{\bit}{\begin{itemize}}
\newcommand{\eit}{\end{itemize}}
\newcommand{\barr}{\begin{array}}
\newcommand{\earr}{\end{array}}
\newcommand{\mm }{\mathrm}
\newcommand{\mcal }{\mathcal}
\begin{document}
\title{Compressed low Mach number flows in astrophysics: a nonlinear
Newtonian numerical solver}

\titlerunning{Low-Mach number flows}

\author{A.~Hujeirat${}^{1}$
        \and F.-K. Thielemann${}^{2}$
        \and J. Dusek${}^{3}$
        \and A. Nusser${}^{4}$}

\authorrunning{A.~Hujeirat et al.}
\institute{
ZAH, Landessternwarte Heidelberg-K\"onigstuhl,\\
Universit\"at Heidelberg, 69120 Heidelberg, Germany \and Fakult\"at
f\"ur theoretische Physik, Universit\"at Basel, Switzerland \and
Insititut de Mecanique et des Solides, Louis Pasteur University,
Strasbourg, France \and Faculty of Physics, Technion, Israel}

\date{Received ... / Accepted ...}

\offprints{A. Hujeirat,  \email{AHujeirat@lsw.uni-heidelberg.de}}

\abstract
  {
Internal flows inside gravitationally stable astrophysical objects,
such as the Sun,  stars and  compact stars are compressed and
extremely subsonic. Such low Mach number flows are usually
encountered when studying for example dynamo action in stars,
planets, the hydro-thermodynamics of X-ray bursts on neutron stars
and dwarf novae. Treating such flows is numerically complicated and
challenging task}
{We aim to present a robust numerical tool that enables modeling the
time-evolution or quasi-stationary of stratified low Mach number
flows under astrophysical conditions.}
{ It is argued that astrophysical low Mach number flows cannot be
considered as an asymptotic limit of incompressible flows, but
rather as highly compressed flows with extremely stiff pressure
terms. Unlike the pseudo-pressure in incompressible fluids, a
Possion-like treatment for the pressure would smooth unnecessarily
the physically induced acoustic perturbations, thereby violating the
conservation character of the compressible equations. \\
Moreover, classical dimensional splitting techniques, such as ADI or
Line-Gauss-Seidel methods are found to be unsuited for modeling
compressible flows with low Mach numbers.}
{In this paper we present a nonlinear Newton-type solver that is
based on the defect-correction iteration procedure and in which the
Approximate Factorization Method (AFM) is used as a preconditioner.
This solver is found to be sufficiently robust and is capable of
capturing stationary solutions for viscous rotating flows with Mach
number as small as $\mcal{M}~\approx ~10^{-3},$ i.e., near the
incompressibility limit.}
   {}

\maketitle
%

\date{Received ... / Accepted ...}


\section{Introduction}

Among different energy contents, the gravitational and thermal
energies in bound astrophysical systems are dominant. The virial
theorem states that in the absence of external pressure and surface
tension the total energy of gravitationally bound system is
negative, i.e.,
 \beq
 -\alpha_1 \DD{G M^2}{R} + 2\left[\mcal{E}_\mm{th} + \mcal{E}_\mm{kin}\right] + \beta_1 \DD{\Phi^2}{R}
 < 0,
 \eeq
where $\alpha_1, \beta_1$ are constants less than one and where

 \beq
   \left\{ \barr{ll}
   {Gravitational~energy}: & \mcal{E}_\mm{grav} = \mm{\DD{GM^2}{R}} \\
   Thermal~energy: & \mcal{E}_\mm{th} = \mm{\DD{3}{2}\int_V P~ dvol} \\
   Kinetic~energy: & \mcal{E}_\mm{kin} = \mm{\DD{1}{2}\int_V \rho |V_f|^2~ dvol} \\
   Magnetic~energy: & \mcal{E}_\mm{mag} = \mm{\DD{{\Phi}^2}{R}},
  \earr
\right . \eeq where $\mm{G,~\Phi, ~M,~R, ~P,~V_f}, $ denote
respectively the gravitational constant, the magnetic flux, the mass
and radius of the object, pressure and fluid-velocity, and $~dvol$
is an infinitesimal volume-element.

The final stage in the evolution of such gravitationally stable
systems is characterized by the following energy measure:
 \beq
|\mcal{E}_\mm{grav}|\geq |\mcal{E}_\mm{th}| \gg
|\mcal{E}_\mm{kin}|,|\mcal{E}_\mm{mag}|. \eeq In terms of velocities
per mass this relation is equivalent to:
 \beq \rm{V^2_g \geq V^2_S
>> V^2_f, V^2_A}, \eeq where the velocities correspond to the
self-gravitating energy $(\rm{V^2_g}\doteq \DD{G M^2}{R})$, thermal,
fluid and magnetic (\Alfven) velocities.

Therefore, fluid motions in gravitationally stable astrophysical
systems are naturally sub-sonic, hence the Mach number is relatively
low.

For example,  helioseismology measurements have revealed that the
Sun oscillates on various frequencies. In particular, it has been
found that the origin of the 5-minute oscillations is a self-excited
sound wave travelling back-and forthwards through the Sun interior
\citep{Musman1974}. This  corresponds roughly to the sound speed:

\beq V_\mm{S} \sim \DD{R_\odot}{\mm{5~minutes}}\approx~2.3\times
10^{8}~\mm{cm~s^{-1}}.
 \eeq

\citet{Roth2002} have suggested that internal flows can have a
maximum sectorial amplitude of about $10^3~\mm{cm~s^{-1}}$. They
argue that a higher velocity would lead to a noticable distortion of
the rotation rate in the convection zone, hence contradicts
observations. In this case, the Mach number reads:  \beq
{\mathcal{M}}= \DD{V_\mm{HD}}{V_\mm{S}} \sim 10^{-4}.\eeq
Consequently, the fluid motions in the Sun is compressible with
extremely low Mach numbers.

Similarly, in the case of neutron stars, the temperature of the
superfluid ranges between $10^7$ up to $5\times10^8$ K, depending on
the crust heat source \citep{VanRiper1991}. The superfluid velocity
 relative to coordinates rotating with angular velocity
$\Omega_\mm{NS}$ can reach  ${\rm {V_\mm{HD}}\approx
10^4-10^6~~cm~s^{-1}}$  \citep{Jones2003}.\\
Thus, the ratio of the sound crossing time to the hydrodynamical
time scale reads: \beq \DD{\tau_\mm{S}}{\tau_\mm{HD}}\approx
\left(\DD{V_\mm{HD}}{V_\mm{S}}\right)^2 = {\mathcal{M}}^2 \approx
10^{-6}, \eeq
  where $V^2_\mm{S} [= dP/d\rho = (\rho - \DD{1}{3}p)/(\rho+p)]$
  corresponds to the the sound speed squared, which is roughly $10\%$
  the speed of light, depending on the equation of state.

The flows in these two extreme astrophysical objects indicate that
numerical solvers should be robust enough to deal with extremely low
Mach number flows. Such flow-conditions are encountered when trying
to model the origin of the solar dynamo or the thermonuclear
ignition of hydrogen rich matter on the surface of neutron stars,
considered to be responsible for Type-I X-ray bursts
\citep{Fisker2005} or for novae eruption in the case of  white
dwarfs \citep{Camenzind2007}.

\section{Compressible versus weakly  and strongly incompressible flows}

While the equations describing compressible and incompressible flows
are apparently similar, the underlying physics and the corresponding
numerical treatments are fundamentally different.

In general, compressible flows are made of plasmas. The internal
macroscopic motions may become either supersonic or extremely
subsonic. Incompressible flows however, are generally made of
liquid, so that a further compression would not lead to a noticeable
change of their density. The transition from gas phase into fluid
phase mostly does not occur via smooth change of the equation of
state. For example, a high pressure acting onto a container of hot
water vapor cannot be asymptotically extended to describe the
pressure in normal water fluid. Therefore, from the astrophysical
point of view, weakly incompressible flows can be viewed as strongly
compressed plasmas, in which the macroscopic velocities are
relatively small compared to the sound velocity.

To clarify these differences, we write the set of hydrodynamical
equations in non-dimensional form using the scaling variables listed
in Table (\ref{Define_Sacling}).

\begin{table}[htb]
\begin{tabular}{cll}
Scaling&variables &  neutron star(interior)
\\\hline
$\tilde{L}$          &Length            &$\sim~ 10^{6}~\mm{cm}$ \\
$\tilde{\rho}$       &Density           &$\sim~10^{14}\rm{~g~cm^{-3}}$\\
$\tilde{\mathcal T}$ &Temperature       &$\sim~10^7$ K\\
$\tilde{\mathcal P}$ &Pressure          &$\sim~10^{26}~\mm{dyn~cm^{-2}}$\\
$\tilde{V}$          &Velocity          &$\sim~10^{6}~\rm{cm~s}^{-1}$\\
$\tilde{B}$          &Magnetic Fields   &$\sim~10^9~ G$  \\
$\tilde{\mathcal M}$ &Mass              &$\sim~M_\odot$\\
 \end{tabular}
\caption{Scaling variables for non-dimensioning the hydrodynamical
equations.}\label{Define_Sacling}
\end{table}
The set of hydrodynamical equations describing compressible plasmas
in conservative form reads: \bit
\item Continuity equation:
\beq
   \DD{\partial \rho}{\partial t} +   \nabla \cdot \rho V = 0,
\eeq
\item The momentum  equations:
\beq
     \DD{\partial \rho V}{\partial t} +  \nabla  \cdot (\rho V\otimes
     V)
    = -\DD{1}{\mathcal{M}^2}\nabla P + \DD{\rho}{\mm{Fr}^2}\nabla \Psi
      + \left(\DD{\mathcal{M}_\mm{mag}}{\mathcal{M}}\right)^2{\nabla\times B\times B}
      + \DD{1}{\mm{Re}}\nabla\cdot \sigma,
\eeq where $\sigma (= \eta(\nabla V + (\nabla
V)^\mm{T})-\DD{2}{3}\eta(\nabla \cdot
  V)I),~\eta=\rho\nu~\mm{~and~},$  $\nabla \Psi$ are the Reynolds stress
  tensor, the dynamical viscosity coefficient and the gradient of the potential energy,
   respectively.
\item The total energy equation:
\beq
     \DD{\partial \mathcal{E}}{\partial t}
     +  \nabla\cdot(\mathcal{E}+p)V
    = \mm{(\DD{\mcal{M}}{Fr})^2}~\rho \nabla\Psi\cdot V
    + \mm{(\DD{\mcal{M}}{Re})^2}~\nabla\cdot (V\sigma) + \mm{\DD{1}{Pe}}~\nabla \cdot (\nu_\mm{T} \nabla T),
    \eeq
    where $\mathcal{E} = \rho (\varepsilon + \DD{1}{2} \mm{V}^2)~$
    and $\nu_\mm{T}$ is the heat diffusion coefficient.\\
    We may simplify the total energy equation by separating the
    internal energy from the mechanical energy and assuming a
    perfect
    conservation of the latter. Hence, we are left with an equation that describes the
    time-evolution of the internal energy:
 \beq
       \DD{\partial \mathcal{E}^\mm{d}}{\partial t}
     +  \nabla\cdot \mathcal{E}^\mm{d}V
    = -(\gamma-1)\mathcal{E}^\mm{d}\nabla\cdot V + (\gamma-1) \{\left(\DD{\mathcal{M}}{\mm{Re}}\right)^2\Upsilon
      +  \DD{1}{\mm{Pe}}\nabla \cdot (\nu_\mm{T} \nabla T)\},
    \eeq
    where $\Upsilon (\doteq \eta |\nabla\cdot V|^2)$ is the dissipation function.
\item  Magnetic equation\\
The magnetic induction equation, taking into account transport and
diffusion in non-dimensional form reads:
 \beq
  \DD{\partial B}{\partial t} =  \nabla \times \langle V \times B
  -\mm{\DD{1}{Re^\mm{mag}}~  (\DD{\mcal{M}^\mm{mag}}{\mcal{M}}) }~
  ~ \nabla \times B\rangle. \eeq

 \eit

\begin{table}[htb]
\begin{tabular}{lll}
Name &Symbol &  Definition
\\\hline
Reynolds number              &Re                & $\mm{{\tilde{V}~\tilde{L}}/{\nu}}$\\
Mach number                 &$\mathcal{M}$     &$\mm{\tilde{V}/\tilde{V}_\mm{S}}$ \\
Reynolds number~(magnetic)  &$\mm{Re^\mm{mag}}$      & $\mm{\tilde{V}~\tilde{L}}/{\nu^\mm{mag}}$\\
Mach number~(magnetic)      &$\mathcal{M}^\mm{mag}$  &$\mm{\tilde{V}_\mm{A}/\tilde{V}_\mm{S}}$ \\
Prantl number            &Pr                &$\mm{\nu/\nu_\mm{T}}$\\
Froude number            &Fr                &$(\mm{\tilde{V}/\tilde{V}_g})^2$\\
Peclet number            &Pe                &$\mm{Re\cdot Pr}$\\
 \end{tabular}
\caption{Nondimensional numbers. In this list, the additional
parameters $\nu,~\nu^\mm{mag},~\nu_\mm{T}$ and $\tilde{V_g}$
correspond to hydrodynamical viscosity coefficient, magnetic
diffusivity coefficient, heat diffusion coefficient and to the
effective velocity of the potential energy $\Psi$(i.e.,
$\tilde{V}^2_g= \nabla \tilde{\Psi},$ respectively. }\label{Numbers}
\end{table}

 On the other hand, incompressible flows are described through the
 following set of equations:
  \beq
 \nabla\cdot V = 0.
 \label{divfree}
 \eeq
 \beq
 {\mathcal{M}^2}\,\rho\left [V_t + (V\cdot\nabla)V\right] = -\nabla P
 + \DD{\mathcal{M}^2}{\rm{Fr^2}}\rho\,\nabla \Psi +
\DD{{\mathcal{M}^2}}{\mm{Re}}\nabla\cdot\sigma \label{EqInco2} \eeq

  \beq
       \DD{\partial T}{\partial t}
     +  \nabla\cdot T V
      = (\gamma-1) \{\left(\DD{\mathcal{M}}{\mm{Re}}\right)^2~\Upsilon
      +  \DD{1}{\mm{Pe}}\nabla \cdot ({\nu_\mm{T}} \nabla T)\},
\label{EqInco5}
    \eeq

Despite the apparent similarity, the pressure in compressible flows
has different physical meaning; in the compressible case the
equation of energy influences the momentum equation through the
equation of state, whereas in the incompressible case, the pressure
is just a lagrangian multiplier with no direct physical meaning.
The set of equations of incompressible flows is characterized by the
following two features: \ben
\item The velocity field must not only evolve as described by the momentum equations,
it  should fulfill the  divergence-free condition also.
\item there is no direct equation that describes the time-evolution
of  the pressure. \een

Therefore, we may use the pressure in the momentum equations to form
an equation that enforces the flow to move in such a manner that the
divergence-free condition is always fulfilled, independent of the
constitutive nature of the flow.

This can be achieved by taking the divergence of the momentum
equation above:

\beq
 \nabla\cdot \left[
{\mathcal{M}^2}\,\rho\left [V_t + (V\cdot\nabla)V\right]
 - \DD{\mathcal{M}^2}{\rm{Fr^2}}\rho\,\nabla \Psi -
\DD{{\mathcal{M}^2}}{\mm{Re}}\nabla\cdot\sigma
      \right]
    = -\nabla\cdot\nabla P  = -\Delta P,
\eeq which is equivalent to the following compact form:

\beq \Delta P = RHS. \eeq The right hand
side (RHS) contains the divergence of the other terms of the momentum equation.\\

The strategy of turning $\nabla P$ in the momentum equations into a
Possion equation to achieve divergence-free motions is  the basis of
different variants of the so called projection method, e.g., the
``Semi-Implicit Method for Pressure-Linked Equation" (SIMPLE) and
``Pressure-Implicit with Splitting Operator" \citep[PISO,~~see][for
further details]{Barton1998}. Similarly, the projection method can
be applied also to the induction equation in magnetohydrodynamics.
Here the induction equation is modified to include the gradient of a
scalar function $\Theta$ as follows:

  \beq
  \DD{\partial B}{\partial t} =  \nabla \times \langle V \times B  +
  \cdot\cdot\cdot\rangle
  + \nabla \Theta.
  \eeq
  Taking the divergence of this equation, we obtain:
  \beq
   \Delta \Theta = \nabla\cdot B.
  \eeq
Therefore, this method violates the conservation of the magnetic
flux. The $\nabla \Theta$ is actually a source function for
generating or annihilating magnetic flux, so that generating of
magnetic monopoles from zero magnetic flux cannot be excluded.
Specifically, if $\nabla \Theta= const.,$ then $\Delta \Theta = 0, $
which means that a constant pumping of magnet flux due to numerical
errors cannot be eliminated by applying a Poisson like-operator.

The matrix form of the projection method applied to the momentum
Equation (\ref{EqInco2}) and to the Possion Equation (\ref{divfree})
is as follows:

\beq \left[ \barr{ll}
 J & G \\
 G^* & 0
\earr \right] \left[\barr{c} V^\mm{n+1} \\ P^\mm{n+1}
 \earr
 \right]
= \left[\barr{c} RHS \\ 0
 \earr
 \right],
 \label{IncoMtrix1}
 \eeq
where the coefficient matrices $J = \D L_\mm{m}/\D V,$ $G = \D
L_\mm{m}/\D P,$ $G^* = \D L_\mm{p}/\D V,$ and ``n+1" corresponds to
the values at the new time level. Applying LU-decomposition, the
matrix Equation (\ref{IncoMtrix1}) can be re-written as: \beq
\left[ \barr{lc}
 J & 0 \\
 G^* & -G^* J^{-1}G
\earr \right]
 \left[ \barr{lc}
 I & J^{-1}G \\
 0 & I
\earr \right]
\left[\barr{c} V^\mm{n+1} \\ P^\mm{n+1}
 \earr
 \right]
= \left[\barr{c} RHS \\ 0
 \earr
 \right],
 \eeq

where I denotes the identity matrix. This equation is solved in two
steps:
\beq \barr{l}\rm{I.}~~~~~~
 \left[ \barr{lc}
 J & 0 \\
 G^* & -G^* J^{-1}G
\earr \right] \left[\barr{c} V^\mm{*} \\ P^\mm{*}
 \earr
 \right]
 = \left[\barr{c} RHS \\ 0
 \earr
 \right]\\
 \rm{II.}~~~~~ \left[\barr{lc}
 I & J^{-1}G \\
 0 & I
\earr \right] \left[\barr{c} V^\mm{n+1} \\ P^\mm{n+1}
 \earr
 \right]
 = \left[\barr{c} V^\mm{*} \\ P^\mm{*}
 \earr
 \right]
 \earr
 \eeq

In general the inversion of the Jacobian, J,  is difficult and
costly, and it is therefore suggested to replace it by the
preconditioning $\tilde{A}$. In this case the above-mentioned two
step solution procedure should be reformulated and solved using the
defect-correction iteration procedure:

\beq \barr{l}\rm{I.}~~~~~~
 \left[ \barr{lc}
 \tilde{A} & 0 \\
 G^* & -G^* \tilde{A}^{-1}G
\earr \right] \left[\barr{c} \delta V^\mm{*} \\ \delta P^\mm{*}
 \earr
 \right]
 = \left[\barr{c} d \\ 0
 \earr
 \right]\\
 \rm{II.} ~~~~~ \left[\barr{lc}
 I & \tilde{A}^{-1}G \\
 0 & I
\earr \right] \left[\barr{c} \delta V^\mm{n+1} \\ \delta P^\mm{n+1}
 \earr
 \right]
 = \left[\barr{c} \delta V^\mm{*} \\ \delta P^\mm{*}
 \earr
 \right],
 \earr
 \eeq
where $\mm{\delta P*= P^\mm{*}-P^\mm{n}},$ ~~$\mm{\delta
P^\mm{n+1}=P^\mm{n+1}-P^\mm{*}}$    and $\mm{d = RHS = J\, V^*}.$\\

 Based on an extension of the projection method, the Multiple Pressure Variable
 method
 for modeling weakly incompressible flows has been suggested \citep[MPV, see][]{Munz2003}.
 Following this scenario, it is argued that in the low Mach number regime
 $({\mathcal{M}}<< 1)$, the variables can be expanded in the following manner:
 \beq q = q^{(0)} +
{\mathcal{M}}\,q^{(1)} + {\mathcal{M}}^2\, q^{(2)} +
\cdot\cdot\cdot,\eeq

However,   contrary to the flow conditions in the Sun or violent
fluid  motions on the surface of compact objects powered by
thermonuclear flashes, the MPV method requires  to set the first
leading terms in the expansion  of the pressure  to zero in order to
assure matching of the solutions in the asymptotic limit. Moreover,
the MPV expansion requires that the Mach number be sufficiently
small in order to get an adequate asymptotic limit.

Almegren et al. 2006 studied the time-evolution of an injected heat
bubble in the atmosphere of a neutron star using different types of
numerical approximations aimed at properly treating  incompressible
flows in the weak regime. They find that their suggested low Mach
number approximation performs relatively well compared to pure
incompressible or  anelastic approximations. The strategy relies on
finding an appropriate function $\beta$ of the initial conditions
such that $\nabla\cdot \beta V =0$ is always fulfilled. The leading
terms of the pressure is set to oppose gravity, so that the core
remains in hydrostatic equilibrium.\\
We note however that the violent flow-motions associated with Type-I
X-ray bursts may not necessary be of low Mach number type, so that
finding the appropriate $\beta$ may turn into a difficult analytical
task.
\begin{figure}[htb]
\centering {\hspace*{-0.2cm}
\includegraphics*[width=5.0cm, bb=70 235 388
550,clip]{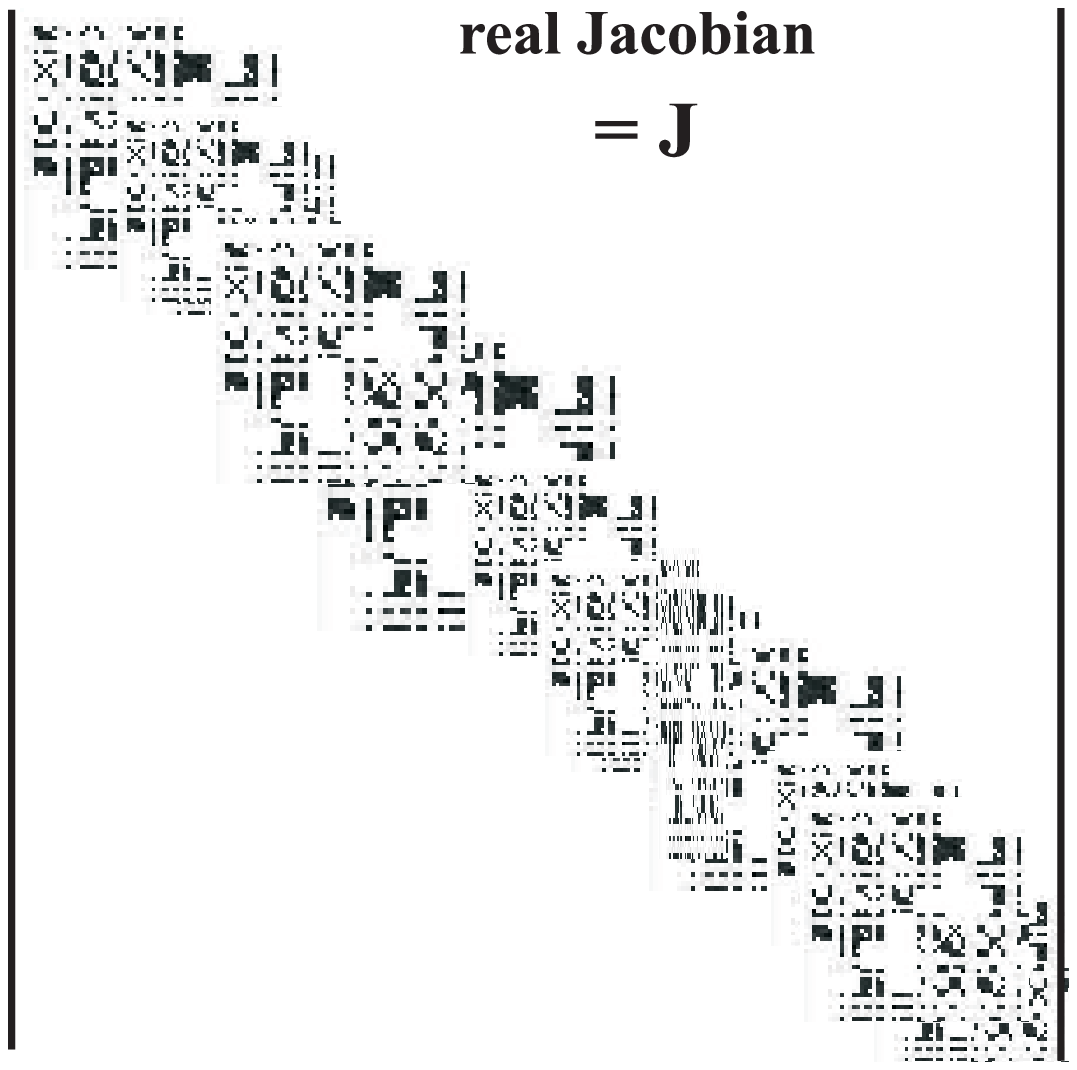}\hspace*{1.0cm}
\includegraphics*[width=5.5cm, bb=35 0 580 520,clip]{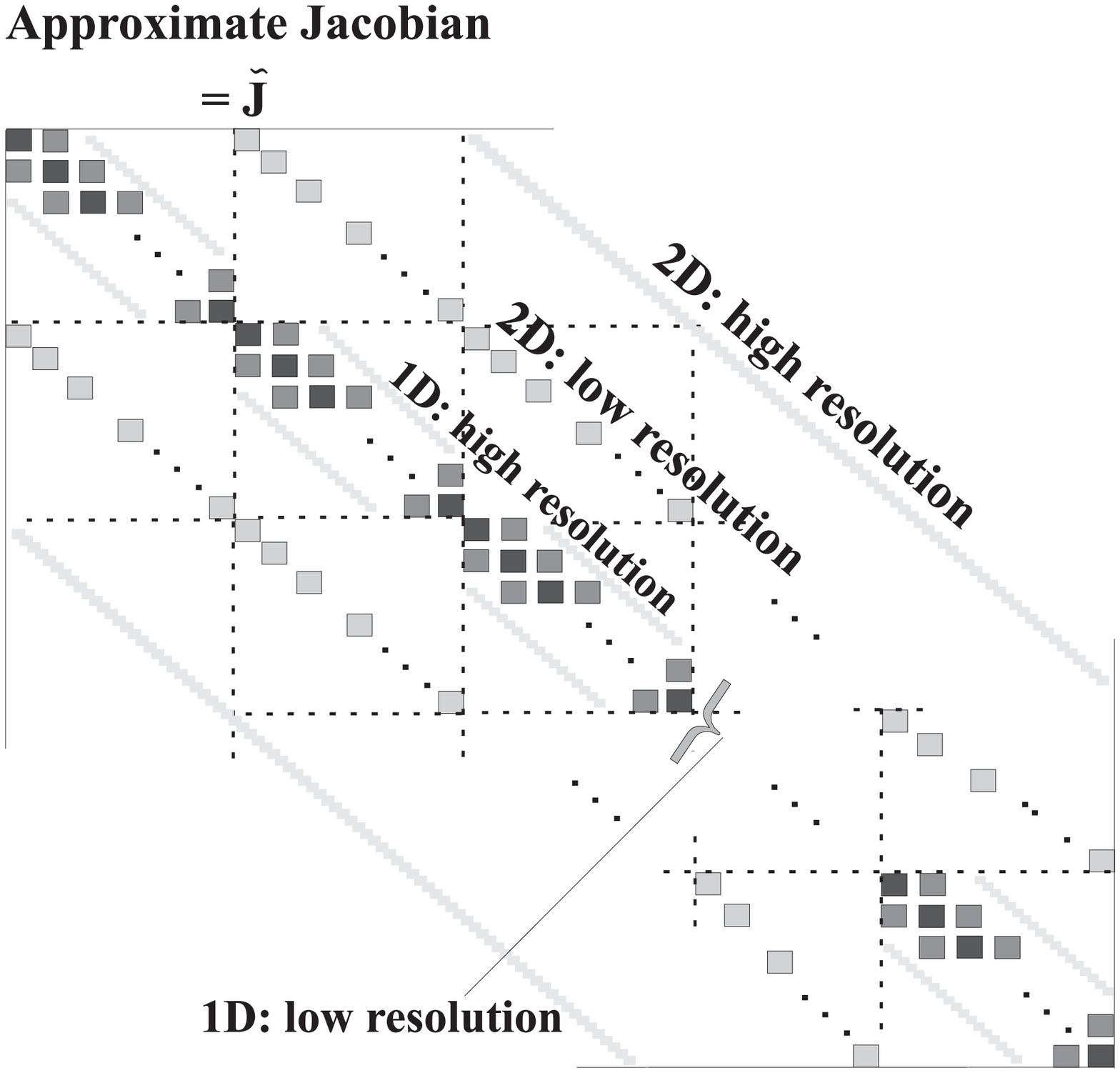} }
\caption{\small  The real Jacobian matrix ($\rm{{J}}$, left) and an
approximate Jacobian matrix ($\rm{\tilde{J}}$, right). In
$\rm{\tilde{J}}$ we have  displayed the entries (blocks) that
correspond to low and high spacial resolution in one and two
dimensions.} \label{Matrix2D}
\end{figure}
\section{Highly compressed low Mach flows: an iterative non-linear
preconditioned Newton solver }
Assume we are given a two-dimensional
nonlinear vector equation of the form:
 \beq \DD{\D q}{\D t} + \DD{\D F(q)}{\D x} + \DD{\D G(q)}{\D y} = f, \label{VEquation1}\eeq
 where $q,F,G,f$ denote the vector of variables, their momentum flux
 both in x and y-directions and a source function, respectively.\\
 Define the residual $\mm{d(q)}$ and look for the vector $\mm{q},$ such that
 $\mm{d(q)}=0$, i.e.,
 \beq d(q) = f - \left[\DD{\D q}{\D t} + \DD{\D F(q)}{\D x} + \DD{\D G(q)}{\D y}\right] =
 0.\eeq
 In the case of a single one-dimensional nonlinear function $\mcal{F}(x)=0$, the zeros can be
 found using the Newton iteration method:
 \beq
 x^\mm{i+1} = x^\mm{i} -
 \DD{\mcal{F}}{\dot{\mcal{F}}},
 \eeq
 where $\dot{\mcal{F}}(x^\mm{i}) = \DD{\D \mcal{F}}{\D
 x}|_\mm{x=x^\mm{i}}$ and ``i" denotes the iteration number.
When applying this approach to a general system of equations such as
Eq. (\ref{VEquation1}), we have then to perform the following
replacements: \beq
   \left. \barr{lll}
   x & \mapsto & q,  \\
   \mcal{F}(x)& \mapsto & d(q) \\
    \dot{\mcal{F}} &  \mapsto & J,
  \earr
\right\}
 \barr{ll}
     &   \\
    \Rightarrow & q^\mm{i+1} = q^\mm{i} - \mm{J^{-1}}{\mm{d}} \\
    &
  \earr
  \label{NIterSystem}
\eeq where J is the Jacobian matrix defined as: $J = \DD{\D R}{\D
q}.$\\
Defining $\mu=q^\mm{i+1} - q^\mm{i}$,  we may re-write Eq.
(\ref{NIterSystem}) as: \beq \rm{J \,\mu = d },\label{JEq}\eeq where
``d" is calculated using arbitrary high spatial and temporal accuracies.\\
The matrix Equation (\ref{JEq}) is said to be:
 \beq \left\{ \barr{lll}
\textrm{Linear\,:}& \mm{if}&  \mm{d=d(q^\mm{n})} \\
\textrm{Otherwise \,:}& \mm{if}&  \mm{d=d(q^\mm{i})}. \\
 \earr
\right. \eeq While in the first case, one need to invert the
Jacobian once per time step, in the second case however, several
iterations per time step might be required to recover the
nonlinearity of the solution. The calculation may become
prohibitively expensive, if the Jacobian to be inverted corresponds
to a system of equation in multi-dimensions to be solved
with high spatial and temporal accuracies. \\
The idea of preconditioning is to calculate the defect ``d" as
proposed by the physical problem (e.g., with very high resolution),
whereas the Jacobian is replaced then by an approximate matrix
$\mm{\tilde{A}}$ of the following properties:
\begin{itemize}
  \item  $\mm{\tilde{A}}$ is easier to invert than J,
  \item $\mm{\tilde{A}}$ and J are similar and share the same
  spectral properties.
\end{itemize}
While the first property is easy to fulfill, the second one is in
general  an effort-demanding issue. It states that the
preconditioning $\mm{\tilde{A}}$ should differ only slightly from
the Jacobian if trivial replacement to be avoided. therefore, given
the matrix $\mm{\tilde{A}}$, the solution procedure would run as
follows:
\begin{enumerate}
  \item Compute the defect ``d".
  \item Use the matrix equation: $\mm{\tilde{A}}\,\mu = \,d.$ to solve for $\mu. $
  \item Update: $q^\mm{i+1} = q^\mm{i} + \mu,$ re-calculate ``d" and
      $\mm{\tilde{A}}$, respectively.
  \item The procedure (2) and (3) should be repeated until $\max(|d|)$
  is smaller than a number $\epsilon$,
  where the maximum runs over all the elements of ``d".
\end{enumerate}

The fundamental question to be addressed still is: how to construct
a robust preconditioner  $\mm{\tilde{A}}$ that is capable of
modeling
low-Mach number flows efficiently, but still easy to invert?\\
In this construction, two essential constrains should be taken into
account:
\begin{itemize}
  \item A conservative  first order spatial discretization of the Navier-Stokes equations generally yields
        a Jacobian matrix of penta-diagonal block form as depicted in Figure (\ref{Matrix2D}).
  \item The gradients of the thermal pressure are dominant, so that the pressure-related terms
  must be
        treated simultaneously.
\end{itemize}
In order to clarify these points, we re-write the matrix Equation
(\ref{JEq}) at an arbitrary grid point (j,k) in the following block
form: \beq
 \begin{array}{lll}
 & {\hspace*{0.3cm}}\overline{S}^\mm{y}_\mm{j,k}{\mu}_\mm{j,k+1} & \\
+ \underline{S}^\mm{x}_\mm{j,k}{\mu}_\mm{j-1,k} & +
{D}^\mm{mod}_\mm{j,k}{\mu}_\mm{j,k} &
  + \overline{S}^\mm{x}_\mm{j,k}{\mu}_\mm{j+1,k}   = d_\mm{j,k} \\
& + \underline{S}^\mm{y}_\mm{j,k}{\mu}_\mm{j,k-1}, &
\end{array}
\eeq where $\underline{S}^\mm{x,y},~\overline{S}^\mm{x,y}$ denote
  the sub and super-diagonal block matrices and
\({D}^\mm{mod} = I/{\delta t} + {D}^\mm{x}+ {D}^\mm{y}\)
the diagonal block matrices,  respectively.\\
While this block structure is best suited for using the one-colored
or multi-colored line Gauss-Seidel iterative method, test
calculations have shown however, that these iterative methods may
stagnate or they may even diverge if the flow is of low mach number
type. The reason for this behaviour is that most iterative methods
rely either on partial updating of the variables or on the
dimensional splitting. These, however, are considered to be
inefficient methods or they may even stagnate, if the system of
equations to be solved are of elliptic type, such as the  Possion
equations.

One way to take the multi-dimensional variations of the pressure all
at one time is to spatial-factorize the Jacobian into sub-matrices,
such that the resulting multiplication results in a good
approximation of the original Jacobian, i.e.,
 \beq
\rm{J\mapsto \Pi_\mm{lm} \tilde{A}_\mm{l}\tilde{A}_\mm{m} }.
  \eeq
The advantages of this method is that the gradients of the pressure
in all direction are incorporated in the matrix $\tilde{A}$.
Furthermore, the right hand side, i.e., the defect is updated only
after all matrices $\tilde{A}_\mm{m}$ are fully inverted. We note
that this preconditioner can be used even as a direct solver as long
as time-dependent solutions are concerned. To clarify this
procedure, we rewrite Equation (\ref{VEquation1}) in the finite
space as follows:
 \beq
 \rm{\DD{\delta q}{\delta t} +
 \DD{\Delta_x F^\mm{n+1}}{\Delta x} + \DD{\Delta_y G^\mm{n+1}}{\Delta y} = 0,}
 \label{VEquation2}\eeq
 where $\rm{\delta q = q^\mm{n+1}-q^\mm{n}}(\mu)$ and $\Delta_\mm{x,y}$ are space difference operators.\\
We may expand $\rm{F^\mm{n+1}}$ and $\rm{G^\mm{n+1}}$ around their
values at the old time levels as follows:
 \[ \mm{F}^\mm{n+1} = \mm{F}^\mm{n}  + {\delta t} \left(\DD{\D F}{\D t}\right)^\mm{n} + \mathcal{O}(\delta t)^2
  = \mm{F}^\mm{n}  + {\delta t} \left(\DD{\D F}{\D q}\right )^n \left(\DD{\D q}{\D t}\right)^\mm{n} + \mathcal{O}(\delta
  t)^2
\mapsto \mm{F}^\mm{n}  + {\delta t} A^n \left(\DD{\delta q}{\delta
t}\right)^\mm{n} + \mathcal{O}(\delta t)^2.
\]
Equivalently,
\[\mm{F}^\mm{n+1}= \mm{F}^\mm{n} + A^n \delta q + \mathcal{O}(\delta
t)^2, \]
 \beq \mm{G}^\mm{n+1}= \mm{G}^\mm{n} + B^n \delta q +
\mathcal{O}(\delta t)^2,  \eeq Substituting these expressions into
Equation (\ref{VEquation2}), we obtain: \beq \rm{\left[
\DD{I}{\delta t} + L_\mm{x}A^{n} + L_\mm{y} B^{n} +
\mathcal{O}(\delta t)^2
 \right]\delta q
 = L_\mm{x}F^{n} + L_\mm{y} G^{n},}
\label{AFM_1} \eeq where $\rm{L_\mm{x},L_\mm{y}}$ denote the
differential operators in x and y-directions, respectively. We may
replace Equation (\ref{AFM_1}) by the following approximation:
 \beq
\rm{\left[ \DD{I}{\delta t} + L_\mm{x}A^{n}\right]\left[I + \delta
t\, L_\mm{y} B^{n}
 \right]\delta q
 = L_\mm{x}F^{n} + L_\mm{y} G^{n},}
 \label{AFM}\eeq
This replacement induces an error which is proportional to : $\delta
t \,L_\mm{x}\,L_\mm{y} + \mathcal{O}(\delta t)^2. $ This error may
diminish for steady conserved fluxes, but may diverge for
time-dependent solutions if the time steps are large. The latter
disadvantage is relaxed by the physical consistency requirement,
that small time steps are to be used if the the sought solutions are time-dependent.  \\
The
matrix Equation (\ref{AFM}) can be re-written in the following
compact form:
 \beq \mm{\tilde{A}_\mm{x}}\,\mm{\tilde{A}_\mm{y}} \mu
= \mm{d},\eeq where $\mm{\tilde{A}_\mm{x}}=  \DD{I}{\delta t} +
  L_\mm{x}A^{n}$ and  $\mm{\tilde{A}_\mm{y}}=  \DD{I}{\delta t} +
  L_\mm{y}B^{n}.$\\

Applying this factorization within a non-linear iterative solution
procedure, the solution procedure would run as follows:

\begin{enumerate}
  \item Compute the defect $\rm{d= d(q^\mm{i},q^\mm{n})}$ at each grid point using the best available
  spatial and temporal accuracies.
  \item Solve the matrix equation: $\mm{\tilde{A}_\mm{x}} \delta q^* = d,$
  to obtain $\delta q^*$.
  \item Solve the matrix equation: $\mm{\tilde{A}_\mm{y}} \delta q = \delta
  q^*$ to obtain $\delta q.$
  \item Update q: $q^\mm{i+1}= q^\mm{i} + \delta q$ and subsequently
  the defect d.
  \item Perform a convergence check to verify if
  $\max(|d|)<\epsilon$. If not, then repeat the algorithmic steps 1-4 repeatedly.
\end{enumerate}
\begin{figure}[htb]
\centering {\hspace*{-0.2cm}
\includegraphics*[width=4.5cm, bb=10 10 375 350,clip]
{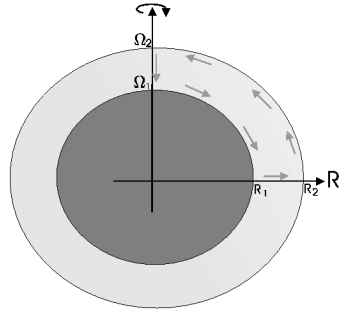} } \caption{\small Two concentric spheres:
the inner sphere has a radius $R_1$ and rotates with angular
velocity $\Omega_1$ whereas the outer one has the radius $R_2$ and
rotates with $\Omega_2$. } \label{2Spheres}
\end{figure}
\section{Taylor-Couette flows between two concentric spheres}
Large scale motions of gas in stellar spherical shells are
controlled by the imbalance of energies, namely between the
potential, thermal, rotational and magnetic energies. It is
generally accepted that rotation deforms surfaces of constant
pressure, but has only indirect influence on surfaces of constant
temperatures. The resulting baroclinicity is unbalanced and derives
large scale meridional circulation \citep{Sweet1950}. On the local
scale, these flows are in general convectively unstable, hence
governed by convective turbulence. Such combined motions are
observationally evident in the solar convective zone.\\
In the laboratory, spherical Couette flows between two rotating
spheres are considered to be similar to rotating stellar envelopes.
In the case of fast rotation, the flow is a combination of primary
azimuthal rotations and a secondary meridional circulation induced
by Ekman pumping \citep{Greenspan1968}. Here the flow is controlled
by two parameters: The Reynolds number and the gap width between the
two spheres. The Reynolds number for this configuration is defined
as: \beq \mm{Re = \DD{|\Delta\Omega|R_1|\Delta R|}{\nu}, } =
\DD{|\Omega_2-\Omega_1|~R_1~|R_2-R_1|}{\nu}, \eeq where $R_{1,2},~
\Omega_{1,2}\mm{~~and ~~} \nu$ are the inner and outer radii, the
angular frequency of the inner and outer sphere and the
viscosity coefficient, respectively (Figure~\ref{2Spheres}).\\
The number of rotationally-induced fluid vortices and transition to
turbulence in Couette flows depend on how large the Reynolds number
is as well on the width $\delta$ of  the gap between the two
spheres. For example, for $\mm{Re> 460}$ and $\delta=1.006$ Couette
flows have been verified to become turbulent
\citep{Gertsenshtein2001}.

\begin{figure}[htb]
\centering {\hspace*{-0.2cm}
\includegraphics*[width=13.5cm, bb=0 175 500 755,clip]
{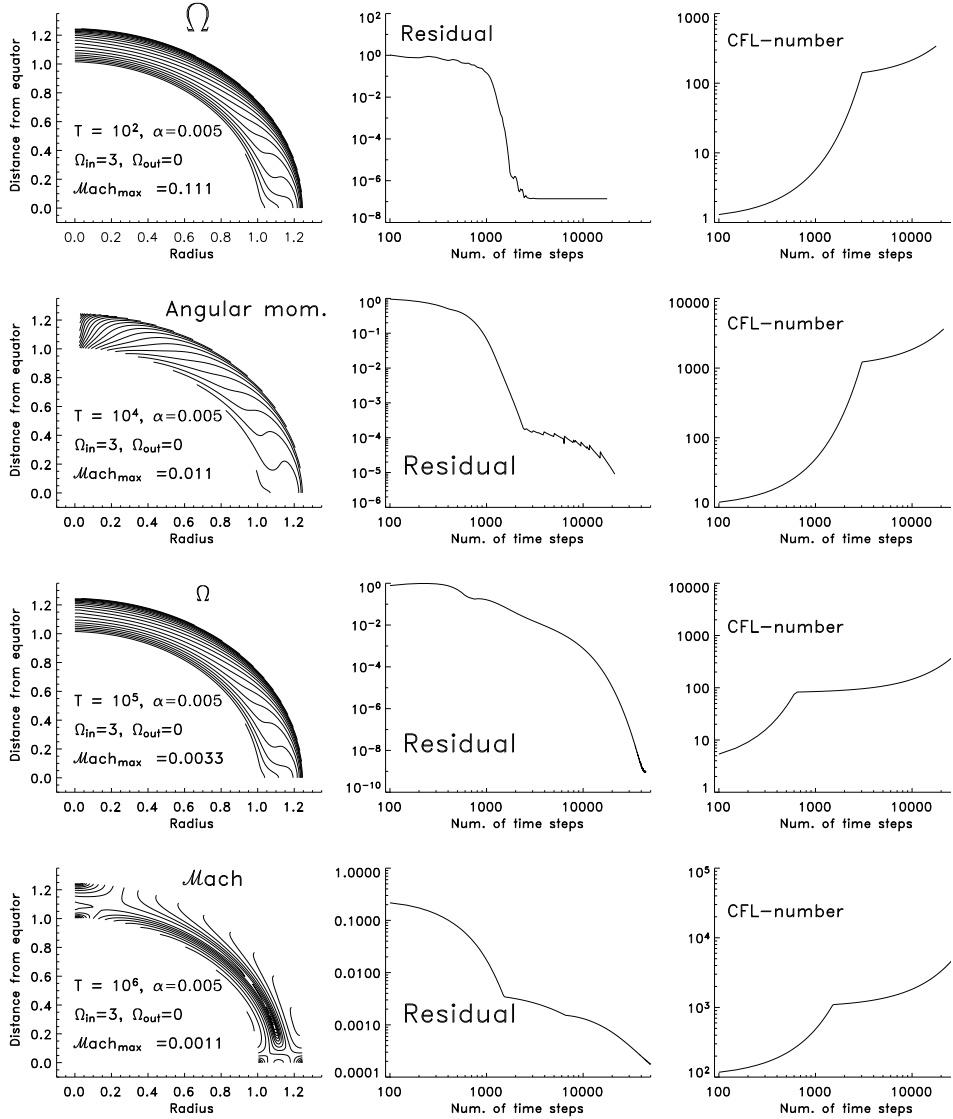} } \caption{\small Taylor-Couette flows: the inner
sphere has a radius $\mm{R_1=1}$ and rotates with $\Omega_1=3$,
whereas
 the outer sphere has a radius $\mm{R_1=1.2}$ and $\Omega_1=0$. The flow has the
  constant viscosity coefficient $\alpha=0.005$.
 The Mach number is set to decrease systematically by
 increasing the temperature from $\mm{T=10^2}$ up to $\mm{T=10^6}$. This corresponds to a reduction of the
 Mach number by two orders of magnitude. The right panel shows the corresponding time-evolution of the time
 step size in units of Courant number.} \label{TaylorFlow}
\end{figure}

In applying our solver to Couette flows between two-concentric
spheres,  the following parameters have been used: $\Omega_1=3,~~
\Omega_2=0, ~~
R_1=1,~~R_1=1.25$ and a viscosity coefficient $\nu=0.005$ (see Figure \ref{TaylorFlow}). \\
The domain of calculation is limited to the first quadrant $[1\leq R
\leq 1.25]\times[0 \leq \theta \leq \pi/2].$ Along the equator and
polar axis reflecting boundary conditions have been imposed, whereas
the outer and inner boundaries are set to be rigid with zero
material flux across them.

In order to test the capability of the solver to deal with low Mach
number flows, we have systematically increased the temperature from
10 up to one million, which corresponds to a reduction of the Mach
number by at least two orders of magnitude.\\
Our results, partially displayed in Figure (\ref{TaylorFlow}),
indicate that AFM as a preconditiong in combination with the
defect-correction Newton  iteration is indeed capable of modelling
weakly incompressible flows down to Mach number $\mcal{M}\approx
10^{-3}$.
%
\section{Summary}
In this paper we have shown that weakly incompressible flows in
astrophysics are actually highly compressible low Mach number flows
in which the pressure plays a vital role in dictating the fluid
motions. Therefore, these flows can be well-treated using a robust
compressible numerical
solver in combination with a sophisticated treatment of the pressure terms.\\
However, as the sound wave crossing time in low Mach number flows is
extremely short relative to the hydrodynamical time scale,
we have concluded that time-explicit methods are not suited.\\
On the other hand, projection methods based on Possion-like solvers
for the pressure violates the conservative formulation of the
hydrodynamical equations, namely the entropy principle and therefore
the monotonicity character of the scheme.

Our conclusion is that the set of hydrodynamical equations
describing the time-evolution of compressible low Mach number flows
should be solved using an implicit robust solver. Our numerical
studies show that nonlinear Newton-type solvers in combination with
the defect-correction iteration procedure and using the Approximate
Factorization Method (AFM) as preconditioning is best suited for
treating such flows. Unlike the classical non-direct methods that
rely on dimensional splitting and/or partial updating, e.g., ADI and
Line-Gauss-Seidel, the AFM is based on factorizing the Jacobian
matrix and subsequently updating the variable in all directions
simultaneously. Similar to the treatments of the pseudo-pressure in
incompressible flows, the pressure gradients in low Mach number
flows do not accept dimensional splitting even when it is applied
within the preconditioning only.
\begin{acknowledgements}
The authors thanks Bernhard Keil for carefully reading the
manuscript.
      This work is supported by the Klaus-Tschira Stiftung under the
project number 00.099.2006.
\end{acknowledgements}

\end{document}